\documentclass{article}
 
\usepackage{bm}
\usepackage{amsmath}
\usepackage{amssymb}
\usepackage{graphicx}
\title{On the classical complexity   of sampling from quantum interference of  indistinguishable bosons   }

\author{V. S. Shchesnovich\\
\small{Centro de Ci\^encias Naturais e Humanas, Universidade Federal do
ABC, Santo Andr\'e,  SP, 09210-170 Brazil }}

\date{\today}
 
\begin{document}
 
\newtheorem{theorem}{Theorem}
\newtheorem{corrolary}{Corollary}

\maketitle

\def\pr{\prime}
\def\be{\begin{equation}}
\def\en#1{\label{#1}\end{equation}}
\def\dag{\dagger}
\def\bar#1{\overline #1}
\def\U{\mathcal{U}}
\newcommand{\per}{\mathrm{per}}
\newcommand{\rd}{\mathrm{d}}
\newcommand{\vare}{\varepsilon }

\newcommand{\bp}{\mathbf{p}}
\newcommand{\m}{\mathbf{m}}
\newcommand{\n}{\mathbf{n}}
\newcommand{\s}{\mathbf{s}}
\newcommand{\bk}{\mathbf{k}}
\newcommand{\bl}{\mathbf{l}}
\newcommand{\br}{\mathbf{r}}

\begin{abstract}
Experimental demonstration of  the quantum advantage over classical simulations  with  Boson Sampling  is currently under   intensive investigation. There seems to be a   scalability       issue to the necessary number of bosons on the linear optical platforms  and  the experiments, such as the recent Boson Sampling with $20$ photons on $60$-port interferometer by H.~Wang~\textit{et al}, \textit{Phys. Rev. Lett.} \textbf{123,}  250503 (2019), are usually carried out on a small  interferometer,   much smaller than  the size necessary for the no-collision regime.  Before   demonstration of quantum  advantage,   it is urgent  to   estimate exactly  how the classical computations  necessary for   sampling from the output distribution of Boson Sampling are reduced when  a smaller-size interferometer is used.  The present work supplies  such a result, valid   with arbitrarily close to $1$ probability, which  reduces  in  the no-collision regime to  the  previous  estimate by  P.~Clifford and R.~Clifford.    One of the results with immediate application to current experiments with Boson Sampling is that   classically sampling from   the interference of $N$ single bosons  on an $M$-port interferometer    is at least as hard as  that     with    $\mathcal{N}= \frac{N}{1+N/M}$ single  bosons in the no-collision regime, i.e., on  a much larger   interferometer with at least  $\mathcal{M}\gg N^2$ ports. 
\end{abstract}

\section{Introduction}
 
 Boson Sampling   idea  of   Aaronson \& Arkhipov \cite{AA}  links  sampling from the output distribution of the many-body   quantum interference of    $N$ indistinguishable single bosons   on  a unitary linear   $M$-port chosen at random from the Haar measure  and    a mathematical problem of estimating matrix permanents of  matrices with elements being independent identically distributed complex Gaussian random variables.   
 The relation is due  the fact that in  the so-called no-collision regime,  with at least  $M\gg N^2$,  when  each  output port receives at most a singe boson  \cite{Bbirthday},  matrix elements of the unitary matrix describing a multiport are approximated by independent identically distributed complex Gaussian random variables.  Under  plausible conjectures,  a  classical simulation  of the above sampling  task to a given   error  $\epsilon$ with the computations      polynomial in  $N$ and  $1/\epsilon$   is impossible \cite{AA}.  The classical hardness originates from the fact, that the amplitudes of  $N$-boson   interferences  are  given as    matrix permanents  of $N$-dimensional submatrices of  a unitary matrix   \cite{C,Scheel},  whose  computation   is believed to be classically hard      \cite{Valiant,A1} (see also a review  \cite{Tutor}).   The fastest known algorithms \cite{Ryser,Glynn}   compute   a matrix permanent of an arbitrary matrix  in $O(N2^N)$  computations.   Even a  relative error approximation  to the absolute value of a   matrix permanent       is classically hard \cite{AA} (given two conjectures are true; see also Ref.  \cite{AntConc}).  Approximation algorithms to an additive error are known: a   probabilistic   approximation to matrix permanent of an arbitrary matrix \cite{Gurv}, generalised for matrices with repeated rows or columns  \cite{DerGurv}   and with repeated rows and columns \cite{UnivBound}.  On the other hand,  deterministic approximation algorithm to a relative error   was found only for    diagonally dominant matrices \cite{ApprPermCMat}.      In contrast,  distinguishable bosons (classical particles)  result in an  output probability distribution  expressed as  the matrix  permanents of positive matrices, which can be estimated  polynomially in $N,1/\epsilon$  \cite{JSV} and  approximated  by  the deterministic  algorithm  of Ref. \cite{DerGurv}. 

Boson Sampling,     is among several   proposals \cite{QSProp,QSArch,QSChar,QSColdAt}   considered for the quantum supremacy demonstration, i.e., demonstration of a provable computational advantage of a quantum system  over digital computers  \cite{Presk},  the  first step on the  way of  using the computational advantage  promised by  quantum mechanics   \cite{F,Sh}. The importance of such a demonstration cannot be  underestimated  in view of   the opposite hypothesis \cite{Kalai}.     

Boson Sampling  can be    easier than  the universal quantum computation \cite{KLM}   implemented  with linear  optics,   since it  requires neither  interaction between photons nor error correction schemes. The  proof of principle experiments    \cite{E1,E2,E3,E4} have followed the initial idea, moreover,      improvements and advances  are  constantly reported  \cite{GBS,E5,TbinBS,TbinBSExp,pure1phBS,LossBS,12phBS}. Recently, an experimental Boson Sampling  with $20$ photons on $60$-port interferometer have been demonstrated \cite{20ph60mod}.   Moreover, alternative  platforms include   ion traps \cite{BSions}, superconducting qubits \cite{BSsuperc},   neutral atoms in optical lattices \cite{BSoptlatt}  and  dynamic   Casimir effect \cite{BScas}. 
  
Before any demonstration of quantum supremacy is attempted,  it is important to know  how exactly the classical computations required to sample from the output distribution of Boson Sampling scale up  with the size of the system, i.e., $N,M$. A significant  reduction in the number of classical computations was demonstrated by using   a Markov Chain  Monte Carlo classical simulation algorithm \cite{QSBS} and subsequent  analytical estimate \cite{Cliffords}  with the      threshold $N$ for the quantum advantage   elevated from the original $N\approx  30$ \cite{AA} to  $N\approx  50$ bosons.      This result is  essentially  the threshold of classical  computability of a single matrix permanent by Glynn's method \cite{Glynn}, which allows  simultaneous computation of matrix permanents appearing in the chain rule of conditional probabilities  with the largest     size of a matrix   equal to  the total number of bosons. 

On the other hand, inevitable experimental  imperfections  can  allow for further speed up,  moreover,  even  efficient classical simulations algorithms  are possible \cite{K1,R1,OB,PRS,RSP}.  Currently used  planar optical  platforms have an exponential scaling   of boson  losses  with the  optical depth   of a multiport   (i.e., the total number of layers of beamsplitters and phase shifters),  leaving only   a finite-size window   in optical depth   for the quantum supremacy demonstration with Boson Sampling  \cite{PRS} and  making  a  large planar optical multiport   $M\gg N^2$  for $N\sim 50$ \cite{QSBS,Cliffords}   unsuitable  for the quantum advantage demonstration due to strong losses.   
 
 One is therefore forced to either  search for other  platforms for experimental  Boson Sampling, which do not  have exponentially scaling losses of bosons, or  consider   sending more photons to an interferometer  up to linear  scaling of the  interferometer size $M\sim N$,  not  discarded as a possibility \cite{AA} for the quantum advantage demonstration.  Current experiments  use   smaller interferometers   than necessary for the no-collision regime  (due to losses or other reasons) \cite{LossBS,12phBS,20ph60mod}.   In such  regimes,  however, bunching of bosons  at output of a unitary multiport    reduces  the  computational hardness  of the corresponding  output probabilities   \cite{Tichy,ComplEST,GenConc},  due to   boson bunching at the output ports    and the fact that the computational hardness of a general matrix permanent  depends on the matrix rank \cite{Barv}.    Thus though the previous algorithm  \cite{QSBS,Cliffords} can be applicable to such a regime, the number of computations can be further reduced due to boson bunching and an appropriate estimates of this effect  is in order.

   To  derive an estimate on  the computational hardness  valid for arbitrary $M\ge N$, or arbitrary  density of bosons   $\rho = {N}/{M}$,   is the goal  of  the present work.     In contrast to the no-collision regime, where  the number of computations for  an output probability  does not depend on the output distribution of bosons, thanks to vanishing probability of boson bunching at the output ports,  for  a finite-density regime $\rho = \Omega (1)$  it depends strongly on the occupations of the output ports \cite{Tichy,ComplEST,GenConc}.    Averaging of the number of computations over the $\binom{M+N-1}{N}$ output distributions of bosons seems to be an unfeasible task, since the output probabilities are hard to estimate.  In an experiment  a fixed  multiport must be   chosen uniformly randomly over the unitary group  (i.e., according to the Haar measure) \cite{AA}, but  averaging over the Haar-random unitary  is not equivalent to  averaging over the output distributions of bosons for  a fixed multiport. Thus, estimating the number of computations for a fixed multiport  for general $M\ge N$ is a difficult problem.     This problem is  solved below by   observing that the   distribution of the total number of output ports occupied by bosons,  averaged over the Haar-random unitary,  is a narrow one of the bell-shaped form.  Since the   tails of such a  distribution  are exponentially small, we cut off two small regions corresponding to  unitary matrices   on the    low and  high ends of the computational complexity, and estimate the number of computations from above and below for the rest of multiports.  In this way, the resulting estimates are valid for a fixed multiport with  a small probability of failure.   This approach allows us to obtain  an experimentally relevant result that applies to a fixed multiport (except a small fraction of multiports  at the low end of the computational complexity, such as the multiports  described by  diagonally dominant unitary matrices \cite{ApprPermCMat}):   simulating Boson Sampling with  $N$ bosons  on $M$-port interferometer   is at least as hard   as  with $\mathcal{N} =N/(1+N/M)$ bosons  in the no-collision regime, i.e., on a $\mathcal{M}$-port  with at least $\mathcal{M}\gg N^2$.  Though such an estimate on the number of computations does not substitute an actual complexity-theoretical proof of hardness of Boson Sampling  in a finite-density regime, it is an important result for the experimental efforts  to scale up Boson Sampling, while simultaneously combat losses of bosons.

 The rest of the text is organized as follows.   In section \ref{sec2}  a modification  of  Glynn's method \cite{Glynn}  is  proposed, with   the same   speed-up  for the matrix permanents of matrices with repeated columns/rows as found previously   in  Ryser's method  \cite{ComplEST}. The classical hardness of a  probability  at the output of Boson Sampling operating in a finite-density regime      depends on the distribution of bosons over the output ports (the output configuration). Therefore, to quantify the classical complexity    of the output probability distribution,  the lower and upper bounds on the number of computations in  the modified Glynn's method are  used.  The  bounds depend on the total number of output ports occupied by bosons.  The crucial fact  is that in a Haar-random multiport the distribution of the total number of ports occupied by bosons  has a bell-shaped form, with the tails bounded by those of a  binomial distribution.   This fact allows to state lower and upper bounds with probability arbitrarily close to $1$ in the Haar measure, theorem 1 of  section \ref{sec3}.  Moreover, since  a variant of   Glynn's method is employed   for computation of matrix permanents,   the  algorithm of Ref. \cite{Cliffords} (applicable uniformly over all regimes of  density of bosons) is used to give a reduced estimate on  the number of computations required to produce a single sample   from the output distribution of  Boson Sampling in any  density  regime $0< \rho\le 1$, section \ref{sec4}, where theorem 2 gives the main result.  In the last section \ref{sec5}  open problems  related to the presented results are discussed. 

\section{Modified   Glynn's formula for  matrix permanent of a matrix  with repeated columns or rows }
\label{sec2}

We consider quantum interference of $N$ perfectly indistinguishable single bosons on a unitary  multiport $U$ with $M$ input and output ports, below fixing  the input ports to be $k= 1,\ldots,N$. We use the notations:     
 $\rho = N/M$ for the density of bosons, 
 \be
 \hat{p}(\m) = \frac{|\per (U[1,\ldots,N|l_1,\ldots,l_N])|^2 }{m_1!\ldots  m_l!}
 \en{Eq1} 
 for the probability \cite{AA} of detecting bosons  in a multi-set of output ports   $l_1\le \ldots\le  l_N$ corresponding to  
 output configuration $\m = (m_1,\ldots,m_M)$, $ m_1+\ldots+m_M = N$,   $U[1,\ldots,N|l_1,\ldots,l_N]$ for  the submatrix of $U$ on the rows $1,\ldots,N$ and columns   $l_1,\ldots, l_N$,  $\per(\ldots)$ for the  matrix permanent   \cite{Minc}. 

  For  a general  multi-set   $l_1, \ldots,  l_N$   with   coinciding  elements, the corresponding  submatrix  \break $U[1,\ldots,N|l_1,\ldots,l_N]$  is  rank-deficient, which reduces the computational complexity of its matrix permanent \cite{Barv}.  An estimate of the   number of classical computations $\mathcal{C}_\m$  for evaluation of such  a matrix permanent  with account of the speed up due to reduced matrix rank   was found  before \cite{ComplEST} by analysing  Ryser's  algorithm \cite{Ryser}.  For $N\le M$   we have 
 \be
\mathcal{C}_\m=  O\left(N^2\prod_{l=1}^M(m_l+1)\right). 
 \en{Est}
 The essential result of Eq. (\ref{Est})  was  reproduced also    in Ref. \cite{GenConc}.  Below, however, we will need  the  number of computations  according to Glynn's algorithm  \cite{Glynn},  since this algorithm is  used  for analysing the sampling complexity  in Ref. \cite{Cliffords}. The algorithm itself is  modified below, in a similar way as in Ref. \cite{DerGurv},  to reduce the number of computations to the bare necessary. 
  
  Let us assume that only $n$  output ports are occupied by bosons and set (without losing the generality)   $m_{n+1} = \ldots m_M=0$. Introducing  for  each $l=1,\ldots, n$ an auxiliary complex variable $x_l$ taking $m_l+1$ values 
\[
x_l  \in\{ 1,  e^{\frac{2i\pi}{m_l+1}}, \ldots, e^{\frac{2i\pi m_l}{m_l+1}}\},
\]
 we can   rewrite the matrix permanent in the output probability $p(\m)$ Eq. (\ref{Eq1}) as follows 
\be
\per (U[1,\ldots N|l_1,\ldots,l_N]) =\frac{\prod_{l=1}^n m_l!}{\prod_{l=1}^n (m_l+1)}\sum_{x_1} \ldots \sum_{x_{n}}x_1\ldots x_{n}\prod_{k=1}^{N} \left( \sum_{l=1}^{n}x_l U_{kl} \right).
\en{Eq2}
Indeed, the r.h.s. of Eq. (\ref{Eq2}) is a sum over   all  permutations   in the  product of $N$  elements of  $U[1,\ldots N|l^\prime_1,\ldots,l^\prime_N]$  with  the  multi-set of columns  $l^\prime_1,\ldots, l^\prime_N$ corresponding to a configuration $\m^\prime = (m^\prime_1,\ldots,m^\prime_n,0,\ldots,0)$,  divided by $m^\prime_1! \ldots m^\prime_n!$, where $m^\prime_l = m_l + \Delta_l(m_l + 1)$ for a whole  number $\Delta_l \ge 0$, since there is at least one factor $x_l$ for all $l=1,\ldots, n$ on the r.h.s of Eq. (\ref{Eq2}) and  each auxiliary variable $x_l$ satisfies 
\[ 
\sum_{x_l}  x^{m^\prime+1}_l = \sum_{j=0}^{m_l} e^{\frac{2i\pi (m^\prime+1)}{m_l+1}j}=\left\{ \begin{array}{cc} 0,& \mathrm{rem}(m^\prime+1,m_l+1)\ne 0\\ m_l+1, &\mathrm{rem}(m^\prime+1,m+1)=0 \end{array}\right.,
\]
where $\mathrm{rem}(s,q)$ is the remainder of division $s/q$. Since  the product of summations over columns of $U$ on  the r.h.s. of Eq. (\ref{Eq2}) contributes  factors consisting of   products  of exactly $N$ elements, i.e., $  x_{l^\prime_1}\ldots x_{l^\prime_N} U_{1l^\prime_1} \ldots U_{Nl^\prime_N}$,  to the  summations over $x_1,\ldots,x_n$,    we must have   
\be
N = \sum_{l=1}^n m^\prime_l = \sum_{l=1}^n \Delta_l(m_l +1) +N,
\en{Eq3}
resulting in  all $\Delta_l=0$, i.e., $m^\prime_l = m_l$. Eq. (\ref{Eq2}) is an alternative generalised Glynn estimator  to that found in Ref. \cite{DerGurv}.

Similar as in  Glynn's formula   \cite{Glynn}, a reduction of the number of summations is still  possible in Eq. (\ref{Eq2}). Assuming that  $m_n$ is the \textit{minimum}   of the non-zero $m_l$  (i.e., $m_n\le m_l$ for $l=1,\ldots, n$) one can set $x_n=1$  and omit  the summation over $x_n$ in Eq. (\ref{Eq2}). Indeed, let us show  that 
\be
\per (U[1,\ldots N|l_1,\ldots,l_N]) =\frac{\prod_{l=1}^n m_l!}{ \prod_{l=1}^{n-1} (m_l+1)}\sum_{x_1}\ldots \sum_{x_{n-1}}x_1\ldots x_{n-1}\prod_{k=1}^{N} \left( \sum_{l=1}^{n-1}x_l U_{kl} + U_{kn}\right).
\en{Eq4}
In this case, due to an  arbitrary number of  factors $U_{kn}$ with different $k$ in the product on the r.h.s. of Eq. (\ref{Eq4}), the equality condition in Eq. (\ref{Eq3}) for $m^\prime_1,\ldots, m^\prime_n$  becomes an inequality  for $m^\prime_1,\ldots, m^\prime_{n-1}$  
\[
N \ge  \sum_{l=1}^{n-1} m^\prime_l = \sum_{l=1}^{n-1} \Delta_l(m_l +1) +N -m_n,
\]
satisfied for $m_l\ge m_n$ only if $\Delta_l=0$ ($m^\prime_l=m_l$)  for all $l=1,\ldots,n-1$. 

Now let us estimate the number of computations in Eq.  (\ref{Eq4}).  There are $N$ multiplications of sums of  matrix elements in the rows $1,\ldots, N$.   The number of  additions  in the inner sum (over $l=1,\ldots,n$)  can be reduced, similarly as in Refs.  \cite{Glynn,Cliffords},     by ordering the  factors  $x_{l_1}\ldots x_{l_{n-1}}$  in the outer sum in such a way that neighbour factors   have just one element $x_{l_j}$ different.   Then, for each such factor, the   computation of the inner sum requires only one addition and one multiplication  (to change one factor $x_{l_j}$) for  each term in the outer sum in Eq. (\ref{Eq4}).  Therefore,  the total number of computations $\mathcal{C}_\m$  in Eq. (\ref{Eq4})   is defined solely  by the outer sum and the product   and reads 
 \be
\mathcal{C}_\m \equiv O\left(N \frac{\prod_{l=1}^n (m_l+1) }{\mathrm{min}(m_l+1)}\right).
\en{Eq5}
Let us note that    Eq. (\ref{Eq5}) correctly estimates  the number of computations $C_\m = O(N)$ in the case of maximally bunched output $\m = (N,0,\ldots,0)$ as well as the correct result for the non-bunched outputs $O(N2^N)$ by the Ryser-Glynn algorithm \cite{Ryser,Glynn}, a  feature not present in any of the  previous estimates \cite{ComplEST,GenConc}. Nevertheless, an interesting  observation is that the reduced estimate  still obeys the majorization pattern   pointed out in  Ref. \cite{MajPatt}. 

 On average, in  a Haar-random multiport $U$,   the speedup  in the number of computations due to Eq. (\ref{Eq4})  (due to the denominator  in   Eq. (\ref{Eq5}))  as compared to Eq. (\ref{Eq2}) is at most  by   $O(\ln N)$.   To show that,  consider the probability of maximal bunching of bosons at the output of a Haar-random network, given as  \cite{Bbirthday}
\be
\mathrm{Prob}( \mathrm{max}(m_l)\le m)  \approx   \left[1-\Bigl(\frac{\rho}{1+\rho}\Bigr)^{m+1}\right]^M . 
\en{Eq6}
Setting in Eq. (\ref{Eq6}) the  probability  to  be  $1-\epsilon$  we  obtain 
\be
m  \lesssim  \frac{\ln\left(\frac{N}{\rho \epsilon}\right)}{\ln\left(\frac{1+\rho}{\rho }\right)}.
\en{mmax}

\section{Classical computations for calculating a single probability of the output distribution of many-boson    interference}
\label{sec3}
 
To  quantify the number of classical computations for general   $ M\ge N$,  when an  arbitrary output configuration $\m=(m_1,\ldots,m_M)$ occurs with a  non-vanishing probability (with the  uniform average probability in a Haar-random network),  we  study how the total number of output ports occupied by bosons is  distributed. It is shown below that, for  $N\gg 1$   in a Haar-random multiport  the total number of  output ports occupied by bosons  has  a  bell-shaped distribution, with the tails bounded by a binomial distribution. This allows one to get an  almost sure  lower and upper bounds on the computational complexity for any network, i.e., the bounds apply with arbitrarily close to $1$ probability to a randomly chosen network according to the Haar measure.  

Given the number $1\le n\le N$ of output ports occupied by bosons, the number of classical computations necessary to compute an output probability $\hat{p}(\m)$, according to the algorithm in Eq. (\ref{Eq4}),  satisfies
\be
\Omega\left(N  2^{n}\right) \le \mathcal{C}_\m \le  O\left(N\left(1+N/n\right)^n \right) ,
\en{Eq7}
where the lower bound is obtained by assuming the maximal bunching in a single output port, e.g., $m_1 = N-n+1$ and $m_l=1$ for $l=2,\ldots,n-1$, whereas the upper bound by uniformly distributing bosons over the occupied output ports in Eq. (\ref{Eq5}), i.e.,  $m_l = N/n$  for $l=1,\ldots,n$ (an upper bound for   $m_l = [N/n]$ for $\le n-1$  and $m_n = N - (n-1)[N/n]$   in Eq. (\ref{Eq5})).

Consider  now the probability distribution of  the number of output ports occupied by bosons in a Haar-random network.  Thanks to the  uniform   average probability $\langle \hat{p}(\m)\rangle = \binom{M+N-1}{M-1}^{-1}$  the probability of $n$ output ports occupied by bosons  is obviously  (here $\theta(x) = 1$ for $x>0$ and $\theta(x)=0$ otherwise)
\be
P(n)\equiv \mathrm{Prob}\left(\sum_{l=1}^M \theta(m_l) = n\right) = \frac{\binom{M}{n}\binom{N-1}{n-1}}{\binom{M+N-1}{M-1}},\quad 1\le n\le N.
\en{Eq8}
Indeed, we choose uniformly randomly  $n$ out of $M$ output ports and   distribute in them $N-n$ bosons with the uniform probability over the output configurations.  The average number $\langle n\rangle$  of the output ports occupied by bosons  reads
\begin{eqnarray}
\label{Eq9}
 \left\langle n \right\rangle  = \frac{MN}{M+N-1} = \frac{N}{1+\rho}\left[ 1 + O\left(\frac{1}{M+N}\right)\right].
 \end{eqnarray}
As $N\gg 1$ (for a fixed $\rho = N/M$) the tails of the  distribution (\ref{Eq8})  (the precise definition of the tails is given below) lie  below  the tails of the following  binomial distribution
\be
B_n(x) =  \binom{M}{n} x^n (1-x)^{M-n},\quad x \equiv   \frac{\rho}{1+\rho}. 
\en{Eq10}
\begin{figure}[ht]
\begin{center}
   \includegraphics[width=.45\textwidth]{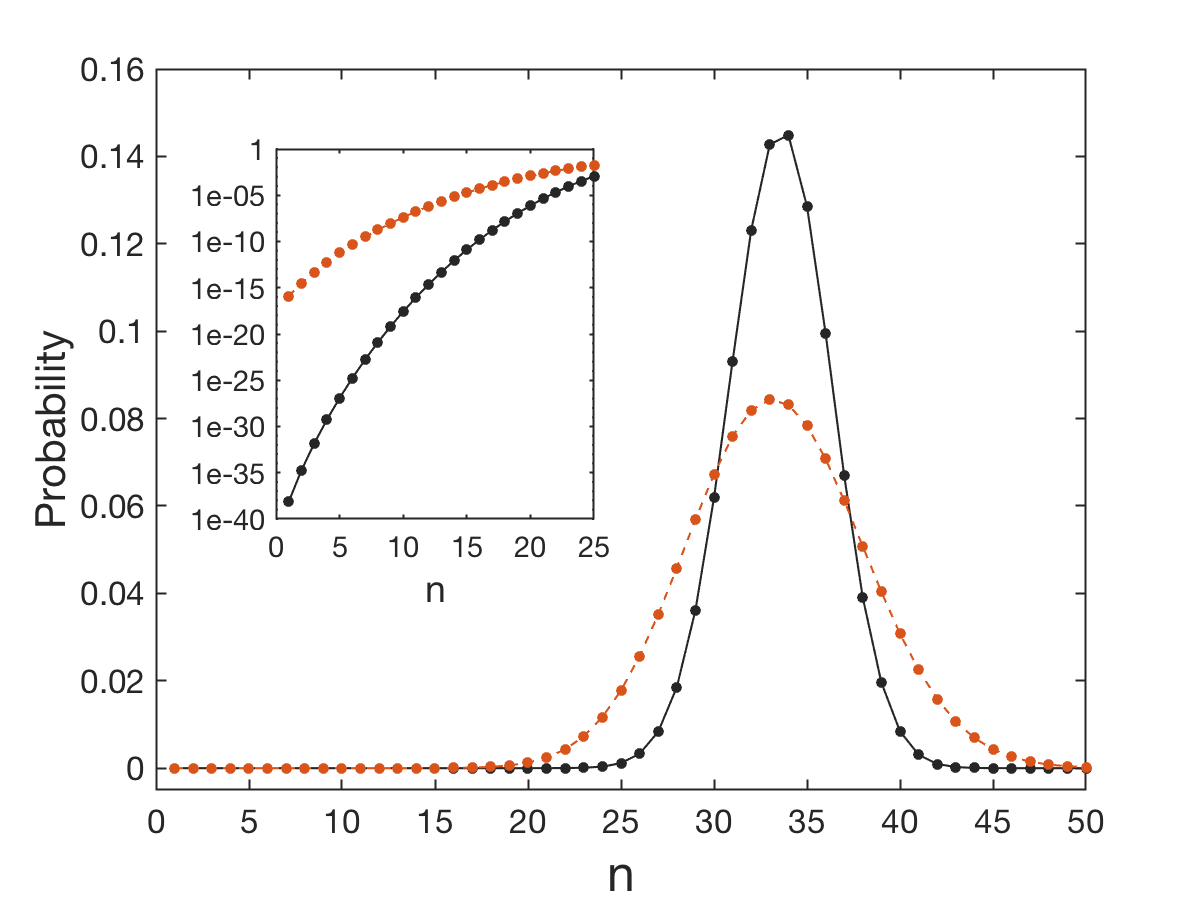}
     \caption{ The main  panel gives the distribution of the number of output ports occupied by bosons Eq. (\ref{Eq8})  (blobs on the solid line) compared to the binomial distribution of Eq. (\ref{Eq10}) (blobs on the dashed line). The insert compares the left tail. Here  $N = 50$, $M=100$. \label{F1} }
   \end{center}
\end{figure}
For $N\gg 1$ the two distributions of Eqs. (\ref{Eq8}) and (\ref{Eq10})  have  bell-shaped form.  Indeed, whereas the binomial is evidently bell-shaped about $n =\overline{ n}  \equiv Mx = N/(1+\rho)$, the distribution of the number of  output ports occupied by bosons satisfies 
\[
P(n+1) =  \frac{M-n}{n+1}\frac{N-n}{n}P(n), \quad P(n-1) =  \frac{n}{M-n+1}\frac{n-1}{N-n+1}P(n),
\]
where the two factors in each of the two relations come from the respective binomial coefficients. Therefore, the only extremum (maximum) probability for $N\gg 1$ is attained  when $n^2 \approx (M-n)(N-n)$, i.e., at the average value of Eq. (\ref{Eq9}). Moreover, by using Stirling's approximation for the factorials \cite{Gosper}, one can easily check that  for $n = N/(1+\rho)$
\[
\frac{\binom{N-1}{n -1}}{\binom{M+N-1}{M-1}} = \sqrt{\frac{1+\rho}{\rho}}\frac{\rho^n}{(1+\rho)^M}\left[1+O\left(\frac{1}{N}\right)\right],\quad x^n(1-x)^{M-n} = \frac{\rho^n}{(1+\rho)^M},
\]
i.e., the maximum of the binomial distribution lies below that of the distribution of the number of output ports occupied by bosons, which fact suggests that at some points $n_\mp$ from the left and from the right of the  point  $ n= \overline{n} $  the binomial distribution dominates that of the total number of occupied ports.

Let us   formally define  the   left  $1\le n\le n_-$ and  right  $n_+\le n\le N$ tails   by  the points  $n_\mp$ of equal probability $P(n_\pm) = B_{n_\pm}(x)$. Now, let us show that such points do exist and are given by the expression $n_\pm = \frac{1\pm \delta_\pm}{1+\rho}N$, for some  $\delta_\pm>0$ when  $N\gg 1$.  To find the points $n_\pm$ consider the equation 
\[
\binom{M}{n}^{-1}P(n) = x^n(1-x)^{M-n}.
\]
Applying   the standard approximation to  factorials \cite{Gosper} in the   binomial $\binom{N-1}{n-1}$,  after simple algebra we obtain  the following asymptotic equation for $\delta_\pm$
 \[
\sqrt{\frac{(1\pm\delta_\pm)(1+\rho)}{\rho \mp \delta_\pm }}\exp\left\{NH\left(\frac{1\pm \delta_\pm}{1+\rho}\right)\right\} = (1+\rho)^N\left[1+O\left(\frac{1}{N}\right)\right],
\]
where $H(z) = -z\ln z - (1-z) \ln(1-z)$, which is asymptotically equivalent (for $\delta_-<1$ and $\delta_+<\rho$) to 
\be
H\left(\frac{1\pm \delta_\pm}{1+\rho}\right) = \ln(1+\rho). 
\en{Eqdelta}
For $0< \rho <   1$ the r.h.s of Eq. (\ref{Eqdelta}) varies between $0$ and $\ln2$, i.e.,  the minimum and maximum values of $H(z)$ for $0\le z\le 1$,  thus there is always a solution for $\delta_\pm>0$ (satisfying also  $\delta_-<1$ and  $\delta_+< \rho$). Therefore, we have shown that there are such $n_\pm = \frac{1\pm \delta}{1+\rho}N$, with some  $\delta= \mathrm{max}(\delta_\pm)\in (0,\rho)$,  that  for $1\le n\le n_-$ and for $n_+\le n\le N$ the binomial distribution of Eq. (\ref{Eq10}) dominates the distribution of Eq. (\ref{Eq8}). 

Now we can use the Hoeffding-Chernoff bound  \cite{Hoeffding},   which states that the tails of the binomial distribution  of Eq. (\ref{Eq10}), i.e., the sum of probabilities for $0\le n\le n_-$ and $n_+\le n \le M$,  with $n_\pm = (1\pm \delta)Mx =  \frac{1\pm \delta}{1+\rho}N$, are    bounded by   $e^{-\delta^2 Mx/4}$.  Therefore, we get for $\delta\in (0,\rho)$
\be
\sum_{n=n_-}^{n_+}P(n)  \ge  \sum_{n=n_-}^{n_+}  B_n(x) >1 - 2 \exp\left(-\frac{\delta^2 N}{4(1+\rho)}\right).
\en{Eq11}
Now,   setting  $ \exp\left(-\frac{\delta^2 N}{4(1+\rho)}\right) =\epsilon/2 $ we get $\delta =  2\sqrt{\frac{1+\rho}{N}\ln (\frac{2}{\epsilon})}$, thus for sufficiently large $N\gg 1$ such $\delta$  will satisfy Eq. (\ref{Eqdelta}) (and $\delta < \rho$, for  a finite-density regime $\rho=\Omega(1)$). Therefore, using  the lower and upper bounds $n=n_\pm=\frac{1\pm \delta}{1+\rho}N$ in Eq. (\ref{Eq7}) we have   the following theorem. 
\begin{theorem}
For any $\epsilon>0$,  with the probability at least $1- \epsilon$ in the Haar measure over the unitary multiports,  the  number of classical computations required to compute (by a modified  Glynn algorithm of section \ref{sec2}) a probability of  the output distribution of quantum  interference of $N$ indistinguishable bosons on a   unitary multiport   with  $M\ge N$ input and output ports satisfies for $N\gg 1$ the following bounds: 
\be
\Omega\left(N2^{\frac{1-\delta}{1+\rho}N}\right) \le \mathcal{C_\m} \le  O\left(N\left(1+r\right)^{\frac{N}{r}}\right), \quad  \delta = 2\sqrt{\frac{1+\rho}{N}\ln\left(\frac{2}{\epsilon}\right)}, \quad r =\mathrm{max}\left(1,\frac{1+\rho}{1+\delta}\right).
\en{Eq12}
\end{theorem}
In formulating theorem 1, by using the max$(\ldots)$ in the definition of $r$ we  have taken into account also the  vanishing density case, and more generally, the case of  $\delta>\rho$, as $N\to \infty$, thus recovering the previous  estimate   $O(N2^N)$ (according to Eq. (\ref{Eqdelta})   the  distribution in Eq. (\ref{Eq8}) does not possess the right  tail in this case).  Moreover, by continuity the validity is extended to the case of $\rho =1$.

 \section{Classical complexity of sampling from  $N$-boson quantum interference on a unitary $M$-port }
 \label{sec4}

  In Ref. \cite{Cliffords} the   estimate $\mathcal{C}=O(N2^N+N^2M)$ on the number of classical computations  $\mathcal{C}$ required to produce a single sample from the output distribution of quantum interference of  $N$ single bosons on a unitary $M$-port was given.  Note that the crucial fact is that the algorithm of Ref. \cite{Cliffords} applies uniformly over all possible output configurations, since  multi-sets of output ports $l_1\le \ldots \le l_N$ are obtained by consecutive  sampling from a conditional probability chain rule used there, each time sampling for one output port.  The leading order in the above estimate is just the number of computations for a single output probability in the distribution. The estimate  of Ref. \cite{Cliffords}, however,    is obtained by using   Glynn's formula \cite{Glynn}, which disregards the speedup  found in  section \ref{sec2},      leading to  the lower and upper bounds  given by theorem 1.   Below we recall the main steps in the algorithm of Ref. \cite{Cliffords} and apply it to derive the sampling complexity with account of the speed up of section \ref{sec3}. 
  
The algorithm of sampling of Ref. \cite{Cliffords} uses symmetry of the output probability Eq. (\ref{Eq1})   with respect to permutations of input and of output ports to enlarge the space of events, given by occupations of the output ports $\m = (m_1,\ldots, m_M)$, $m_1+\ldots +m_M=N$, to the space of independent output ports $l_1, \ldots, l_N$, with $1\le l_j\le M$. This is achieved by using the following summation identity valid  for any symmetric function $f(l_1,\ldots,l_N)$
\be
\sum_\m f(l_1,\ldots,l_N) = \sum_{l_1=1}^M \ldots \sum_{l_N=1}^M \frac{m_1!\ldots m_M!}{N!} f(l_1,\ldots,l_N), 
\en{E1}
where the  summation on the left hand side runs over all $\m$ such that  $m_1+\ldots +m_M=N$. Therefore, by  Eq. (\ref{E1}) the probability $p(l_1,\ldots, l_N)$ of an ordered sequence of  output ports   $l_1, \ldots, l_N$ becomes 
\be
p(l_1,\ldots, l_N) = \frac{m_1!\ldots m_M!}{N!}\hat{p}(\m) = \frac{1}{N!}\left| \per (U[1,\ldots,N|l_1,\ldots,l_N]) \right|^2.
\en{Ep}
Let us now introduce the marginal probability of the first $K$ output ports to be $l_1,\ldots, l_K$ by performing the summation over arbitrary $l_{K+1},\ldots, l_N$:
\be
p(l_1,\ldots,l_K) = \sum_{l_{K+1}=1}^M \ldots \sum_{l_N=1}^M p(l_1,\ldots,l_N) = \frac{1}{N!} \sum_{\sigma,\tau} \left[\prod_{k=1}^K U_{\sigma(k),l_k} U^*_{\tau(k),l_k}\right]\prod_{k=K+1}^N\delta_{\sigma(k),\tau(k)}.
\en{E2}
Observe that, due to the delta-symbols,  the  permutations on the right hand side of Eq. (\ref{E2}) must act as follows $\sigma(1,\ldots, N) = (k_1,\ldots, k_K,k_{K+1},\ldots, k_N)$ and $\tau(1,\ldots, N) = (k^\prime_1,\ldots, k^\prime_K,k_{K+1},\ldots, k_N)$, where $(k^\prime_1,\ldots, k^\prime_K)$ is a permutation of $(k_1,\ldots, k_K)$.  Introducing a permutation $\pi \in S_N$,  which  reorders the input ports in such a way that the first $K$ coincide with $k_1,\ldots, k_K$, expanding  the permutations $\sigma =(\sigma^{(I)} \otimes \sigma^{(II)}) \pi$  and $\tau =(\tau^{(I)} \otimes \sigma^{(II)}) \pi$, with $\sigma^{(I)}, \tau^{(I)}$ acting on the first $K$ and $\sigma^{(II)}$ on the last $N-K$ input ports, and  performing the summation over  $\sigma^{(II)}$  in   Eq. (\ref{E2}) we obtain  the following expression  
\be
p(l_1,\ldots, l_K) = \frac{1}{N!} \sum_{\pi} \frac{\left|\per(U[\pi(1),\ldots,\pi(K)|l_1,\ldots,l_K]) \right|^2}{K!}.
\en{E3} 
In deriving Eq. (\ref{E3}) we have taken into account that for each subset $k_1,\ldots, k_K$ of $K$ input ports there are $K!(N-K)!$ permutations $\pi$ which select the first $K$ input ports from this subset, therefore summation over $\pi$ is accompanied by the factor $K!(N-K)!$ in the denominator with $(N-K)!$ cancelling the same factor in the numerator due to summation over $\sigma^{(II)}$. Now, by  comparing  the right hand side of Eqs. (\ref{Ep}) and (\ref{E3}) one can see that they have a similar form, except the summation over $\pi$ in Eq. (\ref{E3}). The latter summation can be understood as averaging over all possible permutations $\pi\in S_N$, if we assume the uniform probability $p(\pi) = 1/N!$. Under this condition, the marginal probability becomes 
\be
p(l_1,\ldots, l_K) = \sum_\pi p(\pi) p(l_1,\ldots,l_K|\pi), \quad  p(l_1,\ldots,l_K|\pi) \equiv \frac{\left|\per(U[\pi(1),\ldots,\pi(K)|l_1,\ldots,l_K]) \right|^2}{K!}.
\en{E4} 
One can easily verify that our definition of the conditional probability $  p(l_1,\ldots,l_K|\pi)$ in Eq. (\ref{E4}) is self-consistent, i.e., taking a marginal probability of another (larger)  marginal probability results in the marginal probability in the form in Eq. (\ref{E4}). The probability in Eq. (\ref{E2}) can be also put in this form due to the symmetry of the output probability, mentioned above. We have 
\be
p(l_1,\ldots,l_N) = \sum_\pi p(\pi) p(l_1,\ldots, l_N|\pi),
\en{E5}
where there are  exactly   $N!$ coinciding  terms $p(l_1,\ldots, l_N|\pi) =\frac{1}{N!} \left|\per(U[\pi(1),\ldots,\pi(N)|l_1,\ldots,l_N]) \right|^2$.

The sampling algorithm of Ref. \cite{Cliffords} is based on the chain rule for the conditional probability in Eq. (\ref{E4}):
\be
p(l_1,\ldots,l_K|\pi) = p(l_1|\pi) p(l_2|l_1;\pi) \ldots p(l_K|l_1,\ldots,l_{K-1};\pi), \quad p(l_r|l_1,\ldots,l_{r-1};\pi)  \equiv \frac{ p(l_1,\ldots,l_r|\pi)} {p(l_1,\ldots,l_{r-1}|\pi)}.
\en{E6}
Sampling for output ports $l_1, \ldots, l_N$, where $N$ input bosons end up,  is  performed as follows. 
\begin{enumerate}
\item  Sample for permutation $\pi\in S_N$ with the uniform probability $p(\pi) = 1/N!$. This requires $O(N^2)$ computations (e.g., sample uniformly from $1,\ldots, N$ without replacement, which is equivalent to uniformly sampling the permutations by the probability chain rule). 
\item  Given $\pi$ as above, sample for the first output port $l_1$ with the conditional probability $ p(l_1|\pi)$ as in Eq. (\ref{E4}). This requires $O(N)$ computations. 
\item Given the output ports $l_1,\ldots, l_{K-1}$, sample for the next output port $l_K$ by using the conditional probability $p(l_K|l_1,\ldots l_{K-1};\pi)$ of Eq. (\ref{E6}).  Since the previously obtained output  ports are known at this step, the sampling can be achieved by using only the numerator in Eq. (\ref{E6}), i.e., the probability $ p(l_1,\ldots,l_K|\pi)$ for a given set of $l_1,\ldots, l_{K-1}$. Such probabilities (for $K=2, \ldots, N$)  are computed sequentially by using the modified Glynn formula of section \ref{sec2} and the Laplace expansion of the matrix permanent over  unknown output port $l_K$:
\be
p(l_1,\ldots, l_K|\pi) = \frac{1}{K!}\left| \sum_{\alpha=1}^K U_{\pi(\alpha),l_K} \per(U[\pi(1),\ldots, \pi(\alpha-1),\pi(\alpha+1),\ldots, \pi(K)|l_1,\ldots, l_{K-1} ])\right|^2
\en{E7} 
\end{enumerate}

Now, let us show that the  bounds of theorem 1 apply  to the leading order of the number of computations required to produce a single sample  in the above algorithm, if we take into account  Eq.~(\ref{Eq4}).  As above, let  $K$ stand for  the number of sequentially sampled output ports, whereas let   $s$ be the corresponding number of \textit{different}  output ports occupied by $K$ bosons (i.e., $1\le s\le n$, where $n$ is   the total number of  output ports occupied by $N$ bosons, as in section \ref{sec3}). The conditional probabilities are computed by employing    the  Laplace expansion of Eq. (\ref{E7})    in the modified Glynn's formula of Eq. (\ref{Eq4}),  by  following   similar  steps as in  lemma 2 of    Ref. \cite{Cliffords}.    

Consider first the lower bound on computations given by  Eq. (\ref{Eq7}). By Eq. (\ref{E7}) the total number of computations in steps 1 -- 3 above  for cumulative sequences of output ports $l_1,\ldots, l_K$, where $1\le K\le N$,     becomes  
\be 
\mathcal{C}  = O(N^2)+\Omega\left( \sum_{K=1}^N K2^s + MK \right) = \Omega\left( \sum_{s=1}^n\left[ \sum_{K\ge s} K\right]  2^s + MN^2 \right),
\en{Eq13}
where  $\sum\limits_{K\ge s} K$ is the sum over the sequences with $s$ different occupied  ports. The r.h.s. of Eq. (\ref{Eq13})   is obviously  bounded from below by the sequence of sampled ports where the first $N-n+1$ output ports are the same, whereas after this  the sampled sequence gets each time a new  output port.   The  first $(N-n+1)M$ computations of permanents of rank $1$ for different output ports $1\le l\le M$ are dominated by the term $\sum_K MK = MN^2$ in Eq. (\ref{Eq13}). We have therefore $\mathcal{C}\ge \mathcal{C}_{\mathrm{lb}}$ with  
\be
\mathcal{C}_{\mathrm{lb}} =  \Omega\left( MN^2 + \sum_{K=N-n+2}^N K2^{K-N+n-1}\right)= \Omega\left( N2^n +MN^2\right),
\en{Eq14}   
which reduces to the estimate found in Ref. \cite{Cliffords} in the case of vanishing density $\rho \ll 1/N$ (when $n = N$). 
   
Now consider the upper bound on the number of computations $\mathcal{C}^{(K)}_\m$ at each step $K=m_1+\ldots +m_M$.  To derive an upper bound we can  drop the denominator in Eq. (\ref{Eq5}) (see also  the discussion at the end of section \ref{sec2})  to get a simpler upper bound on the number of computations $\mathcal{C}^{(K)}_\m = O\left(N\prod_{l=1}^M(m_l+1)\right)$.  Then  the number of computations $\mathcal{C}^{(K)}_{\m}$ for the steps $K$ and $K+1$ are related as follows  (except a constant factor, omitted for simplicity)
\be
\mathcal{C}^{(K)}_{\m} =  \frac{K}{K+1}\frac{m^{(K)}_l+1}{m^{(K)}_l+2}  \mathcal{C}^{(K+1)}_{\m}\le \frac{K}{K+1}\frac{m+1}{m+2}\mathcal{C}^{(K+1)}_{\m} = \frac{K}{N}\left(\frac{m+1}{m+2}\right)^{N-K}
\mathcal{C}^{(N)}_{\m}, 
\en{Eq14}  
where  $l$ is the output port to which  a boson is added in  step $K+1$ and $m$ is the maximal bunching at an output port, see  Eq. (\ref{mmax}) of section \ref{sec2}. Therefore, with probability  $1-\epsilon$,  the  total number of computations satisfies $\mathcal{C}\le \mathcal{C}_{\mathrm{ub}}$, where
\be
 \mathcal{C}_\mathrm{ub}  = O\left( \sum_{K=1}^N \mathcal{C}_\m^{(K)} + MN^2 \right)  = O\left( (m+2) \mathcal{C}^{(N)}_\m + MN^2\right)
\en{Eq15}
and $m$ is given by Eq. (\ref{mmax}). 
Recalling theorem 1 of section \ref{sec3}, we  have   the following result. 
\begin{theorem}
For any $\epsilon>0$, with the probability at least $1- \epsilon$ in the Haar measure over the unitary multiports, the  number of classical computations $\mathcal{C}$ required to produce  a single sample from the  output distribution of quantum  interference of $N$ indistinguishable bosons on a  unitary multiport with  $M\ge N$ input and output ports satisfies for  $N\gg 1$  the following bounds:
\begin{eqnarray}
\label{Eq16}
&& \Omega\left(N2^{\frac{1-\delta}{1+\rho}N}+MN^2\right) \le \mathcal{C} \le  O\left((m+2)N\left(1+r\right)^{\frac{N}{r}}+MN^2\right),\nonumber\\
&&  \quad  \delta = \sqrt{\frac{4(1+\rho)}{N}\ln\left(\frac{2}{\epsilon}\right)}, \quad r =\mathrm{max}\left(1,\frac{1+\rho}{1+\delta}\right), \quad  m   \lesssim  \frac{\ln\left(\frac{N}{\rho \epsilon}\right)}{\ln\left(\frac{1+\rho}{\rho }\right)}.
\end{eqnarray}
\end{theorem}
Note that, for $N\gg 1$ and a fixed  $ \epsilon$,  $r$  is a growing function of    $\rho$, whereas  $(1+r)^\frac{1}{r}$ is monotonously decreasing with the lower bound  for $r=2$ (for $\rho = 1$ and $\delta =0$) being $\sqrt{3}\approx 1.73$. In its turn, $2^\frac{1-\delta}{1+\rho}$ is also  decreasing with  $\rho$ with the minimum (for $\rho =1$ and $\delta=0$) $\sqrt{2}\approx 1.41$. 

The upper and lower bounds of theorem 2 reduce to the estimate of Ref. \cite{Cliffords},  $\mathcal{C}=O(N2^N+N^2M)$,  in the no-collision (vanishing density) regime, i.e., when $\rho\ll 1/N$. 

From the lower bound in Eq. (\ref{Eq16}) one can conclude that for $N\gg1$ the  $N$-boson quantum interference  in the finite-density regime is at least as  hard to simulate classically as that in the no-collision regime with $\mathcal{N}=N/(1+\rho)$ bosons.

\section{Open problems}
\label{sec5}

In the present work we have focused on single bosons at the input  of a unitary multiport,   however, one can   adopt the results for  arbitrary Fock states at the input. We are   bound by the modified Glynn formula Eq. (\ref{Eq4}) which can use  \textit{either}  repeated columns or rows to obtain a speed up over  the usual Glynn formula \cite{Glynn}.  For  an input  Fock state $\mathbf{s} = (s_1,\ldots, s_k)$, $s_1+\ldots +s_k=N$, one can swap the   input ports and   the  output ports (i.e.,   rows and columns) in   Eq. (\ref{Eq2}) to  obtain a similar   expansion   where now multiple occupations of the input ports$, s_1,\ldots, s_k$ are used instead  $m_1, \ldots, m_n$ of   the output ports. In this way, one obtains an estimate on the number of computations  similar to that of Eq. (\ref{Eq5}) with $s_1,\ldots, s_k$ replacing  $m_1,\ldots, m_n$.  Hence, the    minimum runtime for either column-based  or row-based expansion   reads
\be
\mathcal{C}_{\mathbf{s},\m} = O\left(  N\mathrm{min}\left\{  \frac{\prod_{j=1}^k (s_j+1) }{\mathrm{min}(s_j+1)},  \frac{\prod_{l=1}^n (m_l+1) }{\mathrm{min}(m_l+1)} \right\}\right).
\en{Eq5A} 
It is left as an open  question if  there is   a more advanced  generalisation of Glynn method, or that of Eq. (\ref{Eq4}),  that requires smaller number of computations in  the case of Fock states at the input than the estimate given by Eq. (\ref{Eq5A}).  A related problem   is to find the  speed up of the number of computations to produce a single sample  due to Fock state at the input (multiple occupations of the input ports). Such a speedup can be possible  already with the estimate of Eq. (\ref{Eq5A}). Indeed,  in the algorithm of section \ref{sec4} the permanents  of  sizes $1\le K\le N$ are sequentially computed. For $K\ge 2$ in each such computation one could either employ the row or column based expansion as in Eq. (\ref{Eq2}). Therefore, at each size $K$ the computation could take advantage over the best  speedup due to either repeated rows or columns. This problem    is left for future work.

Another open question is   direct estimate of the tails of the  probability  distribution of output ports occupied by bosons, Eq. (\ref{Eq8}),  instead  of using the binomial distribution which  results in sufficient but not necessary bounds on the tails of the distribution.  With the direct estimate of the tails,  tighter   bounds on  the number of  computations (i.e., smaller gaps between lower and upper bound), than those  in  theorems 1 and 2, could be obtained.  Moreover, this would allow to generalise the  estimate on the  complexity of theorems 1 and  2 of sections \ref{sec3} and \ref{sec4} to  arbitrary (non-fractional) density of bosons   $\rho = N/M$, i.e., for  the case of Fock states input  with   $M< N$. Indeed, to prove the theorems we have used the auxiliary binomial distribution in section \ref{sec3} which bounds the tails of the distribution of the number of ports occupied by bosons, but the important step given by   Eq. (\ref{Eqdelta}), which defines the tails,  has  solutions only for $\rho\le 1$.

 \section{Acknowledgements}  
This work   was supported by the National Council for Scientific and Technological Development (CNPq) of Brazil,  grant  304129/2015-1 and by the S\~ao Paulo Research Foundation (FAPESP), grant 2018/24664-9.


\end{document}